\documentclass[12pt,fleqn]{article}
\usepackage{graphicx}
\usepackage{amsfonts}
\usepackage{amsmath,amssymb,amsthm}
\newcommand{\R}{\mathbb R}
\renewcommand{\r}{\smallsize \mathbb R}

\newcommand{\be}{\begin{equation}}
\newcommand{\ee}{\end{equation}}
\newcommand{\bea}{\begin{eqnarray}}
\newcommand{\eea}{\end{eqnarray}}
\newcommand{\eps}{\epsilon}

\newcommand{\ep}{\qquad {\vrule height 10pt width 8pt depth 0pt}}

\newcommand{\grintl}{[\kern-.18em [}
\newcommand{\grintr}{]\kern-.18em ]}
\newcommand{\trip}{\|\kern-.08em |}
\newcommand{\ds}{\displaystyle}

\newtheorem{theorem}{Theorem}[section]
\newtheorem{thm}[theorem]{Theorem}

\newtheorem{lem}[theorem]{Lemma}
\newtheorem{cor}[theorem]{Corollary}
\newtheorem{prop}[theorem]{Proposition}

\def\smallR{\hbox{\scriptsize I\kern-.23em{R}}}
\def\R{\hbox{$\mit I$\kern-.33em$\mit R$}}
\def\C{\hbox{$\mit I$\kern-.6em$\mit C$}}
\def\un{\hbox{$\mit I$\kern-.77em$\mit I$}}
\def\A{{\boldsymbol{A}}}
\def\p{{\boldsymbol{\cal P}}}
\def\0{\hbox{$\mit I$\kern-.70em$\mit O$}}
\def\r{I\kern-.277em R}

\def\dist{\mbox{\rm dist}}

\newtheorem{assump}{Assumption}

\renewcommand\Im{\mathrm{Im}}
\renewcommand\Re{\mathrm{Re}}

\newcommand\supp{{\rm supp}\,}

\begin{document}

\title{An Adiabatic Theorem for Resonances}

\author{
Alexander Elgart
\thanks{Partially
Supported by National Science Foundation
Grant DMS--0907165.}\\
and\\
George A. Hagedorn
\thanks{Partially
Supported by National Science Foundation
Grants DMS--0600944 and DMS--0907165.}
\\
Department of Mathematics,\qquad and\\
Center for Statistical Mechanics, Mathematical Physics,\\
and Theoretical Chemistry,\\
Virginia Polytechnic Institute and State University,\\
Blacksburg, Virginia~ 24061-0123,~ U.S.A.\\
\vspace{1cm}}

\date{}
\maketitle

\centerline{\it Dedicated to the memory of Pierre Duclos.}

\vskip 15mm
\begin{abstract}
We prove a robust extension of the quantum adiabatic theorem.
The theorem applies to systems that have resonances instead of bound states,
and to systems for which just an approximation to a bound state
is known.
To demonstrate the theorem's usefulness in a concrete situation, we apply
it to shape resonances.
\end{abstract}

\newpage
\baselineskip=20pt

\section{Introduction}
\setcounter{equation}{0}

%
%
%
The goal of this paper is to present a quantum adiabatic theorem
that is general enough to apply to situations in which Hamiltonians
have resonances instead of bound states or if just an approximation
to a bound state is known. The hypotheses of our main
result, Theorem \ref{thm:main}, do not specifically mention
resonances, so we demonstrate how one applies the result by
considering the specific situation of shape resonances.
We plan to apply our theorem to other resonance situations in the future.

Our application to shape resonances has considerable
overlap with the work of Abou--Salem and Fr\"ohlich \cite{ASF},
although many of the details are quite different.
In some instances, we obtain sharper estimates.

\vskip 5mm
The adiabatic theorem of quantum mechanics describes the long time
behavior of solutions to the time--dependent Schr\"odinger equation
when the Hamiltonian generating the evolution depends slowly on time.
The theorem relates these solutions to spectral information of the
instantaneous Hamiltonian.

The traditional quantum adiabatic theorem applies to Hamiltonians that
have an eigenvalue which is separated from the rest of the
spectrum by a gap. Some more recent versions do not require
the gap condition. All one really
needs for the adiabatic theorem is a spectral
projection for the Hamiltonian that depends smoothly on time. This
allows situations where an eigenvalue is embedded
in the absolutely continuous spectrum of the Hamiltonian.
Since embedded eigenvalues are intrinsically
unstable, they usually become resonances once the system is
perturbed. It is intuitively clear that on the time scales which are
shorter than the resonance lifetime, there should not be much of
the difference between a proper bound state and the corresponding
resonance. This intuition leads to the question whether an
adiabatic theorem holds if there is a
{\it nearly} spectral projection for the Hamiltonian
that depends smoothly on time. The main abstract theorem of this paper,
Theorem \ref{thm:main}, provides an affirmative answer to this question.

The paper is organized as follows:~
We state our abstract result in Section 1.1. We then
describe its application to shape resonances in Section 1.2.
Section 2 contains the proofs.~
Technical details are collected in the Appendix.

%
%
%
\subsection{The Extended Adiabatic Theorem}
To state an adiabatic theorem precisely, it
is convenient to replace the physical time $t$ by
the rescaled time $s=\epsilon\,t$. One is then concerned with the
solution of the initial value problem
\begin{equation}\label{SCHReqn}
i\,\epsilon\,\dot\psi_\epsilon(s)\ =\ H(s)\ \psi_\epsilon(s),\qquad
\mbox{with}\qquad\psi_\epsilon(0)\ = \ \psi_0,
\end{equation}
for  small values of $\epsilon$. The Hamiltonian $H(s)$ is
self--adjoint for each $s$ and depends sufficiently smoothly on $s$
in an appropriate sense, and $\psi_\epsilon$ takes values in the
Hilbert space. We shall be more specific about what we mean by
smoothness below. Typically, $s$ is kept in a fixed interval, so that
the physical time $t$ belongs to an interval of length $O(\eps^{-1})$.

\par Our main result, Theorem \ref{thm:main}, hinges on
three assumptions given below. The first is a static condition.
The second is a dynamic condition imposed on the family
$H(s)$. The third controls the relative boundness of the
rate at which $H(s)$ changes.

\vskip 5mm\noindent
{\bf Definition}\quad
We say that a projection $P(s)$ is~
{\it nearly spectral}~ for $H(s)$ if it is self-adjoint and
\be\label{eq:cond_ns}
\left\|\,\left(H(s)\,-\,E(s)\right)\ P(s)\,\right\|\ \le\
\delta/2,
\ee
for all $s\in[0,\,1]$, where $\delta$ is a small parameter.

\vskip 5mm\noindent {\bf Definition}\quad We say that a
projection $P(s)$ is~ {\it smoothly nearly spectral}~ for $H(s)$ if
it is self-adjoint,~ $P(0)$ is nearly spectral for $H(0)$,~ and
\be\label{eq:cond_sns}
\left\|\,\frac{d\phantom{i}}{ds}\,
\Big\{(H(s)\,-\,E(s))\ P(s)\Big\}\,\right\|\
\le\ \delta/2,
\ee for all $s\in[0,\,1]$.

\vskip 5mm\noindent
{\bf Remarks}
\begin{enumerate}
\item[{\bf 1.~}] If $P(s)$ is smoothly nearly spectral,
then it is nearly spectral.

\item[{\bf 2.~}] If $P(s)$ is the spectral projection for energy $E(s)$,
then it is smoothly nearly spectral with $\delta=0$.
\end{enumerate}

\vskip 5mm
\begin{assump}\label{assump_1}
~We assume that there exists either a smoothly nearly spectral or a
nearly spectral projection $P(s)$ for $H(s)$.
\end{assump}

\newpage\noindent
{\bf Remarks}
\begin{enumerate}
\item[{\bf 1.~}]
As we have already mentioned, it is reasonable to expect adiabatic
behavior of the system whenever $\eps$ is small,
but $\epsilon\gg\delta$.
\item[{\bf 2.~}]
From the Weyl Criterion (Theorem VII.12 of \cite{RS-I}), a non-trivial
nearly spectral projection exists for any point $E$ in the spectrum of
a self-adjoint operator. However, we shall impose further dynamical
assumptions on $P(s)$ below. In general, the dynamical assumptions
limit the set of suitable $E(s)$ to eigenvalues or resonances.
\item[{\bf 3.~}]
The notion of a nearly spectral projection is also related to the ideas
of Spectral Concentration. See, {\it e.g.}, Section XII.5 of \cite{RS-IV}.
\end{enumerate}

\vskip 5mm
Assumption \ref{assump_1} is a static condition imposed on the
family $H(s)$. Our results require a dynamic hypothesis as well:
We let $g_a$ be a smoothed characteristic function that takes
the value $1$ at $0$.~ More precisely, we assume
\be\label{eq:g_a}
g_a(x)=1
\quad\mbox{if}\quad |x|<a/2,\qquad
g_a(x)=0
\quad\mbox{if}\quad|x|>a,\quad\ \mbox{and}\quad\ g_a\in C^4(\R).
\ee

\vskip 5mm
\begin{assump}\label{assump_2}
~There exists $a\in(0,\,1)$ such that
\begin{equation}\label{eq:cond_g_2}
\left\|\,
g_a\Big(H(s)\,-\,E(s)\Big)\ \dot P(s)\ P(s)
\,\right\|\
\le \ \delta',
\end{equation}
uniformly for $s\in[0,\,1]$.
\end{assump}

\vskip 5mm
\noindent {\bf Remarks}
\begin{enumerate}
\item[{\bf 1.~}]
Here~ $\delta'$~ is another small parameter, while~ $a$~
should be thought of as an auxiliary tuning
mechanism, which eventually could be optimized. In our
application to shape resonances~ $\delta'$~ is roughly of the
same order of magnitude as~ $\delta$.
\item[{\bf 2.~}]
Intuitively, condition \eqref{eq:cond_g_2} quantifies the
rate at which the wave function ``leaks'' from the range of $P(s)$
to energetically close states.
One can also think of Assumption \ref{assump_2} as requiring a
bound on the spectral density in the vicinity of the resonance
or bound state. Indeed, if there are no bound states with energies
close to that of the resonance, the expression in \eqref{eq:cond_g_2}
tends to zero as~ $a$~ tends to zero.
The second possibility is that there are bound states nearby ({\it e.g.},
pure point spectrum), but their overlap with $\dot P(s)$ is small.
The latter occurs in the Anderson localization problem.
\item[{\bf 3.~}]
We remarked earlier that non-trivial nearly spectral projections exist
for any self-adjoint operator. However, if the operator depends on
a parameter $s$, it may be impossible to construct a corresponding
family of nearly spectral
projection $P(s)$ that satisfies Assumption \ref{assump_2}. Theorem
\ref{thm:sr} shows that both Assumptions \ref{assump_1} and
\ref{assump_2} can be satisfied for shape resonances.
\end{enumerate}

\vskip 5mm
Finally, because we are dealing with unbounded perturbations, we
impose the following technical requirement:
\begin{assump}\label{assump_3}
~We assume that the operators $H(s)$ are self-adjoint on a common
domain ${\cal D}$, and that the derivatives $\dot H(s)$ and $\ddot H(s)$
exist as operators from ${\cal D}$ to the Hilbert space ${\cal H}$.
Furthermore, we assume there exists $C$, such that for all $s\in[0,\,1]$,
$$
\left\|\,\dot H(s)\ \left(H(s)\,-\,i\right)^{-1}\,\right\|\
\le\ C\,\qquad\mbox{~and~}\qquad
\left\|\,\ddot H(s)\ \left(H(s)\,-\,i\right)^{-1}\,\right\|\
\le\ C.
$$
\end{assump}

\vskip 5mm\noindent
Here and throughout the text, $C$ stands for a generic constant.

\vskip 5mm
Our main abstract theorem is the following:
%
%
%
\begin{thm}\label{thm:main}
~Suppose $P(s)$ is a nearly spectral projection
for the Hamiltonian $H(s)$ that satisfies Assumptions
\ref{assump_1} -- \ref{assump_3}.~
Then there exists a unitary propagator for (\ref{SCHReqn}).

Let~ $\psi_\epsilon(0)\in {\rm Range}\,P(0)$.\\[2mm]
If $P(s)$ is nearly spectral, then for all $s\in[0,1]$,
\begin{equation}\label{eq:main}
\hspace{-1cm}
{\rm dist}\left\{\,\psi_\epsilon(s),\ {\rm Range}\,P(s)\,\right\}\
\le\ 2\ \delta'\ +\ 2\ \|\dot P(s)\|\ \trip g_a\trip_3\,\delta\ +\
K_a\,\epsilon\ +\ \delta/\epsilon\,.
\end{equation}
If $P(s)$ is smoothly nearly spectral, then for all $s\in[0,1]$,
\begin{equation}\label{eq:main_1}
\hspace{-1cm}
{\rm dist}\left\{\,\psi_\epsilon(s),\ {\rm Range}\,P(s)\,\right\}\
\le\ 2\ \delta'\ +\ 2\ \|\dot P(s)\|\ \trip g_a\trip_3\,\delta \ +\
\frac{C\,\delta}{a^2}\ +\ C\,\tilde K_a\,\epsilon\ +\
\delta/\epsilon\,.
\end{equation}
The $a$--dependent
constants\, $ K_a$ and $\tilde K_a$ in these expressions are given
in \eqref{eq:K_a} (respectively \eqref{eq:K_a'}) below. The norm
$\trip g_a\trip_3$ is one of the norms described in
\cite{Davies} and Appendix B of \cite{hs}. These norms have the form
\[
\trip g_a\trip_{n+2}\ =\
C\ \sum_{k=0}^{n+2}\ \left\|g_a^{(k)}\right\|_{k-n-1},\qquad
\mbox{with}\qquad
\|f\|_l\ =\ \int\ (1+x^2)^{l/2}\ |f(x)|\ dx,
\]
where the $C$ depends only on $n$.
\end{thm}

\noindent
{\bf Remarks}
\begin{enumerate}
\item[{\bf 1.~}] If $P(s)$ is a spectral projection, we can take
$\delta=0$, and \eqref{eq:main_1} corresponds to the (slightly
improved) adiabatic theorem of \cite{ae}. Even in the case of a
bound state it is often technically simpler to construct an
approximant $P(s)$ for the bound state projection rather than the
exact eigenprojection.
The error in the approximation then contains the parameter $\delta$.

\item[{\bf 2.~}] If $P(s)$ corresponds to an eigenprojection for an
isolated eigenvalue of $H(s)$, then it is typically relatively easy to
verify its smoothness (with respect to the $s$ variable), as it is
``inherited'' from the smoothness of $H(s)$.
Otherwise, checking this is usually highly non-trivial ({\it c.f.}
\cite{GH} where this task is carried for the ground state of the atom
in a QED picture). One of the main obstacles in the implementation of
the quantum adiabatic algorithm \cite{FGGS} is controlling the
norms of the derivatives of $P(s)$. From this point of view,
\eqref{eq:main_1} is a better result than \eqref{eq:main}, as it
only requires a bound on the first derivative of $P(s)$. We note
that even if $P(s)$ is only nearly spectral, one can get a bound in
the adiabatic theorem that depends only on the $L^1$ norm of
$\|\dot P(s)\|$,
but not on $\ddot P(s)$. That can be achieved using a mollifier
argument. In our application to shape
resonances, we have the good control on the smoothness of
$P(s)$. (See Lemma \ref{lem:closedness}.)
\end{enumerate}

\vskip 1cm
%
%
%
\subsection{The Application to Shape Resonances}
Theorem \ref{thm:main} applies to Schr\"odinger operators with
shape resonances. The analysis is not much more complicated for
certain magnetic fields, so we include them.
We consider Schr\"odinger operators~
$H(s)\,:=\,\p_\A\cdot \p_\A\,+\,V(s)$~ on~ $\R^d$,~
where~ $\p_\A = -i\,\hbar\,{\bf\nabla} - \A(x)$.

\begin{assump}\label{assump_VP}
We assume the components of the vector potential~ $\A$~
and their first derivatives~ $\partial_i\A$~ for~ $i=1,...,d$~
are bounded on $\R^d$.
\end{assump}

Shape resonances (see, {\it e.g}. \cite{cdks}) are resonances of
$H(s)$ that arise because the particle can be confined
to a region of space that is bounded by a
classically forbidden region.
If the resonance has energy near $E$, then
we define the classically forbidden region to be
\[
J\ :=\ \{\,x\in \R^d:~ V(x)\,>\,E\,+\,b\,\}
\]
for some $b>0$. One usually assumes that $J$ separates
$\R^d$ into a bounded interior and an unbounded exterior component.
The intuition is that the particle spends a long time in the interior
component, but can eventually tunnel to the exterior component.

We examine this situation where the energy $E(s)$, potential $V(x,\,s)$,
and classically forbidden region $J(s)$ all depend on $s$.

For simplicity, we assume that for an
appropriate value of $E(s)$, $J(s)$ separates $\R^d$ into an
exterior region $O(s)$ and a single connected interior region $I(s)$,
so that
\[
\R^d\ = \ J(s)\,\bigcup\,O(s)\,\bigcup\,I(s).
\]
(See Figures 1 and 2 below.)
%

\vskip -5mm
{\includegraphics[height=2.25in,width=5in]{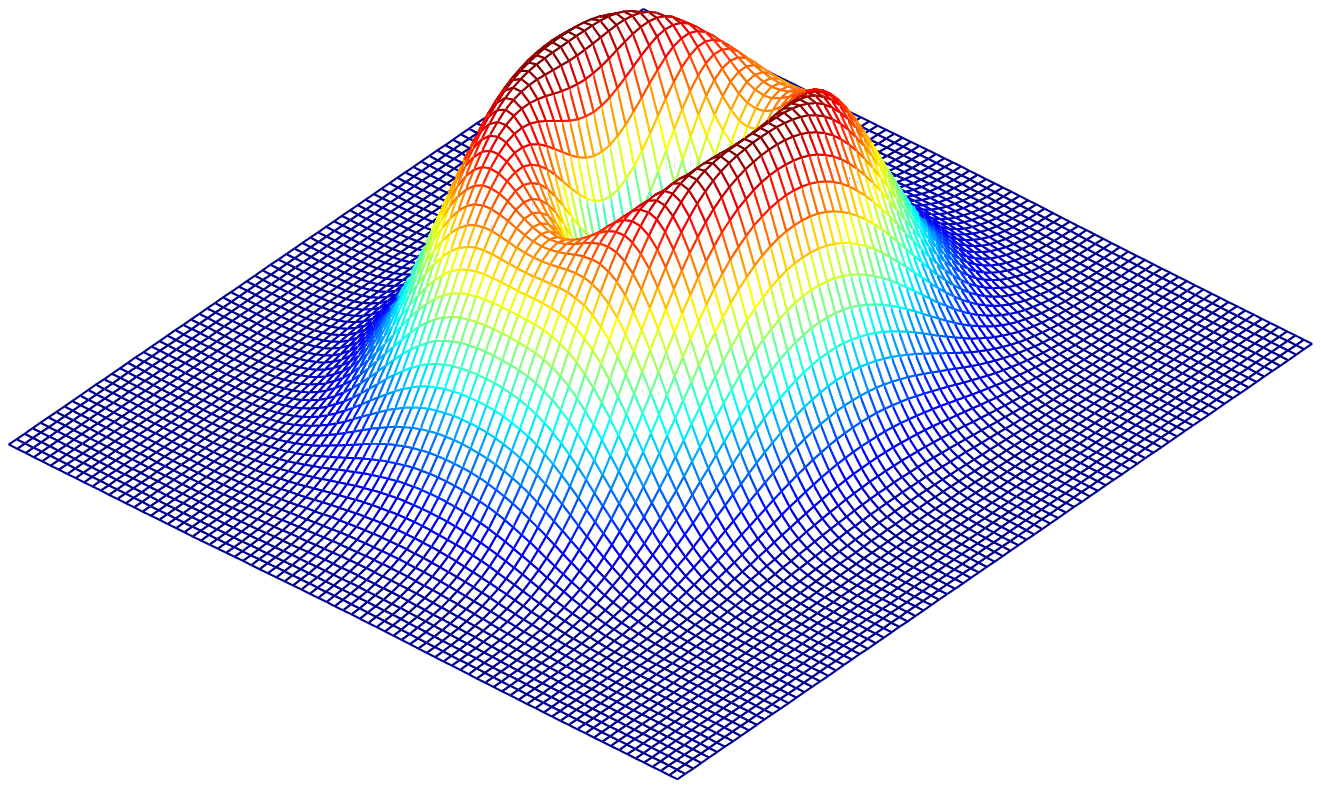}}

\vskip 4mm
\noindent {\bf Figure 1.} A two dimensional potential that has
shape resonances.

\vskip 4mm
\centerline{\includegraphics[height=3in,width=3in]{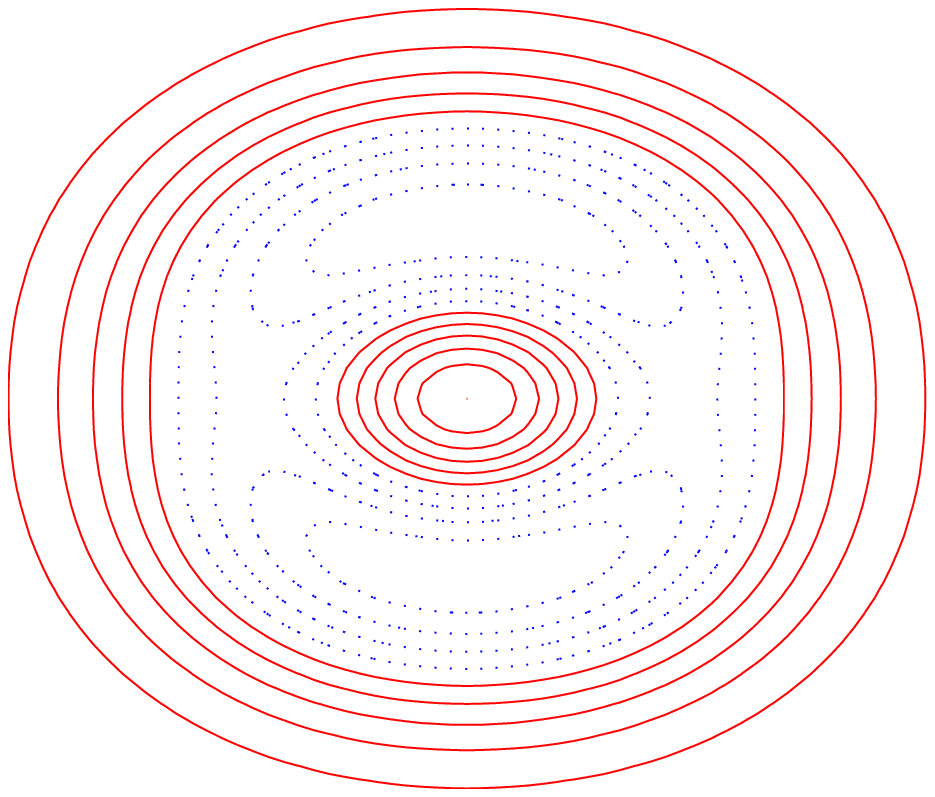}}

\vskip 4mm
\noindent {\bf Figure 2.} Contour plot for the same potential
showing the classically allowed regions (red, solid curves) and
classically forbidden region (blue, dotted curves) for a particular
energy.

%
%

\vskip 8mm
We assume that $H(s)$ has the following
properties for each $s\in[0,\,1]$:

\begin{assump}\label{assump_sr}
\begin{enumerate}
\item[ ]
\item For every $s\in[0,\,1]$, we assume $H(s)$ is a self-adjoint
operator on an $s$--independent domain ${\cal D}$.
\item
For every $s\in[0,\,1]$, there exist $c>0$ and an open set
$\Omega(s)\subset\R^d$, such that
$$
{\rm dist}\left\{\,O(s),\ \Omega(s)\,\right\}\ \ge\ c,\qquad\mbox{and}\qquad
{\rm dist}\left\{I(s),\ \Omega^c(s)\,\right\}\ \ge\ c.
$$
Moreover, the Friedrichs extenstion $H_\Omega(s)$ of the Dirichlet restriction
of $H(s)$ to $\Omega(s)$ is bounded from below.
\item The operator $H_\Omega(s)$ has a discrete eigenvalue
$E(s)$ that depends smoothly on $s$.
\item
$E(s)$ is separated by a spectral gap~ $\Delta(s)>\Delta>0$~ from the
rest of the spectrum of $H_\Omega(s)$.~ For convenience,
we assume~ $\Delta<1$.
\item The potential $V(s)$ satisfies\quad
$\ds \left\|\,V_J(s)\,\right\|\ \le\ C$,\\[4mm]
$\ds
\left\|\,\dot V(s)\ \left(H_\Omega(s)-i\right)^{-1}\,\right\|
\ \le\ C$,
\qquad\mbox{and~}\qquad
$\ds\left\|\,\ddot V(s)\ \left(H_\Omega(s)-i\right)^{-1}\,\right\|\
\ \le\ C\,$,\\[4mm]
where $V_J$ stands for the Dirichlet
restriction to the set $J$.\\
We further assume that~
$\ds{\rm supp}\ \dot V(s)\,\subset\,I(s)\,$.
\end{enumerate}
\end{assump}

\noindent
{\bf Remarks}
\begin{enumerate}
\item[{\bf 1.~}] In typical situations, part 4 of this assumption forces
$\Delta$ to be $O(\hbar)$.
\item[{\bf 2.~}] Assumption \ref{assump_sr} implies a ``Combes--Thomas''
estimate for $H_{J(s)}(s)$.
The (improved) Combes--Thomas estimate \cite{GK} is usually stated
for the operators acting on the whole space $\R^d$.
The extension to the sub-domain case is presented as Theorem
\ref{thm:CT} in the Appendix.
\end{enumerate}
\vskip 5mm \noindent
The following theorem and its corollary are our
main results concerning shape resonances.
%
%
%
\begin{thm}\label{thm:sr}
~Suppose that Assumption \ref{assump_sr} is satisfied.
Then there exists a family $P(s)$ of
projections that satisfies
Assumptions \ref{assump_1}\,--\,\ref{assump_3}
for energy $E(s)$.
The values of the corresponding parameters are
\begin{equation}\label{eq:par_val}
\delta\ =\ C\ \hbar\ e^{-\eta/\hbar},\qquad
a\ =\ \Delta/2,
\qquad\mbox{and}\qquad
\delta'\ =\ C\ \hbar\
e^{-\eta/\hbar}\ \trip g_a\trip_{4}/\Delta\,,
\end{equation}
where $\eta>0$.
\end{thm}
Combining Theorems \ref{thm:main} and \ref{thm:sr} we obtain
%
%
%
\begin{cor}
~In the context of a shape resonance, the adiabatic theorem holds
for\\
$\ds\epsilon\,\gg\,\hbar^{-3}\,e^{-\eta/\hbar}$,~ where $\eta>0$.~
The error term in the adiabatic theorem is bounded by a constant times
$$
\hbar\ \Delta^{-4}\ e^{-\eta/\hbar}\ +\
\hbar\ \Delta^{-3}\ e^{-\eta/\hbar}\ +\
\Delta^{-4}\ \epsilon\ +\
\epsilon^{-1}\ \hbar\ e^{-\eta/\hbar}.
$$
\end{cor}

\vskip 4mm
\noindent
{\bf Proof}\quad
The only nontrivial part of the proof of the corollary is estimating
factors in (\ref{eq:main}) that come from $\dot P(s)$ and $\ddot P(s)$,
some of which occur in $K_a$. However, $P(s)$ from Theorem \ref{thm:sr}
is constructed in Section \ref{subsect2.2} from $P_\Omega(s)$.

We begin by estimating derivatives of $P_\Omega(s)$ by writing
$$
P_\Omega(s)\ =\ -\ \frac 1{2\pi i}\
\oint_{|z-E(s)|=\Delta/2}\ (H_\Omega(s)-z)^{-1}\ dz.
$$
The length of the contour is~ $\pi\,\Delta$,~ while self-adjointness and
the gap condition assure that the norm of the integrand is bounded by~
$2/\Delta$. 

The derivative is
$$
\dot P_\Omega(s)\ =\ -\ \frac 1{2\pi i}\
\oint_{|z-E(s)|=\Delta/2}\ (H_\Omega(s)-z)^{-1}\
\dot V(s)\ (H_\Omega(s)-z)^{-1}\ dz.
$$
By Assumption \ref{assump_sr} and the first resolvent formula, the norm
of the integrand is bounded by $C/\Delta^2$,~ so~
$\|\dot P_\Omega(s)\|\le C\pi/\Delta$.~
Similarly,~ $\|\ddot P_\Omega(s)\|\le C\pi/\Delta^2$.

The corollary is an immediate consequence of these estimates
and Lemma \ref{lem:closedness}.
\hfill\ep

\vskip 4mm
\noindent
{\bf Remarks}
\begin{enumerate}
\item[{\bf 1.~}] As remarked earlier, $\Delta$ is typically bounded above and
below by constants times $\hbar$.
When this is the case, the error in the adiabatic theorem is bounded by a
constant times
$$
\hbar^{-3}\ e^{-\eta/\hbar}\ +\
\hbar^{-4}\ \epsilon\ +\
\epsilon^{-1}\ \hbar\ e^{-\eta/\hbar}
$$
for small $\hbar$.
\item[{\bf 2.~}] There are situations in which eigenvalues of $H_{\Omega}$ can cross
as $\hbar$ is decreased. In those situations, we obtain no error estimate.
\item[{\bf 3.~}]
As we have already commented, the Weyl Criterion guarantees existence of nearly
spectral projections at any point of the spectrum of a quantum Hamiltonian.
However, for general Schr\"odinger operators,
the ranges of those projections contain only functions are very delocalized
in space. The nearly spectral projections we construct for shape resonances
are localized in a region that can be chosen independent of $\delta$ and $\hbar$.
\item[{\bf 4.~}]
There are numerous definitions of resonances
for a quantum mechanical Hamiltonian $H$. One is as follows:~
Since $H$ is self-adjoint, the complex valued function
\[
f(z)\ :=\ \langle\,\psi,\,(H-z)^{-1}\,\psi\,\rangle
\]
is analytic in the upper half-plane for any vector $\psi$.
For $\psi$ in some appropriate set, $f$ has an
analytic continuation $\tilde f$ to some portion of the lower
half-plane. A pole of $\tilde f$ at $E-i\Gamma$ corresponds to
a resonance near energy $E$ with lifetime proprotional to $1/\Gamma$.
This definition is often not practically tractable.
\end{enumerate}

%
%

\section{Proofs}
\setcounter{equation}{0}

%
%

\subsection{Proof of Theorem \ref{thm:main}}
Using Assumption \ref{assump_3}, we obtain existence of the
unitary propagator $U_\eps(s)$
by applying Theorem X.70 of \cite{RS-II}.

Without loss of generality we can assume that~ $E(s)\equiv 0$~
throughout the proof. Indeed, the dynamics generated by~ $H(s)$~
differs from the one generated by~ $H(s)-E(s)$~ only by the
(dynamical) phase\quad
$\exp\left(\,-\,i\,\int_0^s\,E(t)\,dt/\epsilon\,\right)$.

The first step in many proofs of the adiabatic theorem
({\it e.g.}, \cite{ae}) is the construction of the so-called
adiabatic evolution. This is a unitary family $U_a(s)$, such that
\begin{equation}\label{eq:adia}
P(s)\ = \ U_a(s)\ P(0)\ U_a(s)^*\,.
\end{equation}
The second step of the proof is to verify that $U_a$ stays close
to the true evolution $U_\epsilon$, determined by
\begin{equation}\label{eq:true}
i\ \epsilon\ \dot U_\epsilon(s)\ =\ H(s)\ U_\epsilon(s),\quad
\mbox{with}\quad
U_\epsilon(0)\ =\ I.
\end{equation}

We follow this strategy with one modification: Since
$P(s)$ is {\it not}\, a spectral projection, it is hard to construct
an evolution that satisfies (\ref{eq:adia}) and is close to
$U_\epsilon$. Instead, we construct a {\it nearly}\, adiabatic
evolution $U_n(s)$ and replace (\ref{eq:adia}) by
\begin{equation}\label{eq:n-adia}
\hspace{-1cm}
\left\|\,P(s)\ U_n(s)\ - \ U_n(s)\ P(0)\,\right\|\ \,=\ \,
\left\|\,P(s)\ - \ U_n(s)\ P(0)\ U_n(s)^*\,\right\|\ \, \le\ \,
\delta/\epsilon.
\end{equation}
With this modification, the second step is the same as
in the traditional adiabatic theorem.

Specifically, we let $U_n(s)$ be the solution to the initial
value problem
\begin{equation}\label{eq:near_ev}
i\ \epsilon\ \dot U_n(s)\ =\ H_n(s)\ U_n(s),
\qquad\mbox{with}\qquad
U_n(0)\ = \ I\,,
\end{equation}
where the generator is~
$H_n(s)\ =\ H(s)\ +\ i\ \epsilon\ [\dot P(s),\,P(s)]$.
Existence of $U_n(s)$ is guaranteed
by Theorem X.70 of \cite{RS-II}.\\[4mm]
In order to check that the estimate (\ref{eq:n-adia}) holds, we
compute
\begin{equation}\label{eq:der}
\frac{d\phantom{i}}{ds}\ \bigg(\,U_n(s)^*\ P(s)\ U_n(s)\,\bigg)\ = \
\frac{i}{\epsilon}\ U_n(s)^*\ [H(s),\,P(s)]\ U_n(s),
\end{equation}
where we have used
\[
\dot P(s)\ =\ \dot P(s)\ P(s)\ +\ P(s)\ \dot P(s)\qquad\mbox{and}\qquad
P(s)\ \dot P(s)\ P(s)\ =\ 0.
\]

The condition (\ref{eq:cond_ns}) and the unitarity of $U_n(s)$
guarantee that the right hand side of (\ref{eq:der}) is bounded in norm by
$\delta/\epsilon$. We integrate both sides of (\ref{eq:der}) and use
the fundamental theorem of calculus to see that
\be\label{2.16.5}
\|\,U_n(s)^*\ P(s)\ U_n(s)\ -\ P(0)\,\|\ \le\
\delta/\epsilon,
\ee
for all $s\in[0,\,1]$. The estimate
(\ref{eq:n-adia}) follows from the unitarity of $U_n$.

Next we show that the physical evolution $U_\epsilon$ stays close to
$U_n$ for all $s\in[0,\,1]$.

We claim that for a nearly spectral projection,
\bea\nonumber
\|\,U_n(s)\ -\ U_\eps(s)\,\|&=&
\|\,U_\epsilon(s)^*\ U_n(s)\ -\ I\,\|
\\[3mm]\label{eq:compari}
&\le&
2\ \delta'\ +\ 2\ \|\dot P(s)\|\ \trip g_a\trip_3\ \delta\
+\ K_a\ \epsilon,
\eea
where
\bea\nonumber
K_a&:=&2\ \trip g_a\trip_3\ \max_r\,\|\dot P(r)\|^2
\ +\ 2\ \trip g_a\trip_3\ \max_r\,\|\dot P(r)\|
\\[3mm]\label{eq:K_a}
&&+\quad C\ \left(\,\trip g_a\trip_3\ \max_r\,\|\ddot P(r)\|
\,+\,\trip g_a\trip_4\ \max_r\,\|\dot P(r)\|\,\right). \eea

For a smoothly nearly spectral projection, we can improve this bound to
\be\label{eq:compari'}
\|\,U_n(s)\ -\ U_\eps(s)\,\|\ \,\le\ \,
2\ \delta'\ +\ 2\ \|\dot P(s)\|\ \trip g_a\trip_3\
\delta \ +\ C\,\delta/a^2\ +\ C\,\tilde K_a\ \epsilon,
\ee
where
\bea\nonumber\label{eq:K_a'}
\tilde K_a&:=& \max_s\left\{\trip g_a\trip_5\ \
+\ \trip g_a\trip_4\ \|\dot P(s)\|\ + \
\frac{1}{a^2}\left(1\,+\,\|\dot P(s)\|\right)\right\}\,.
\eea

To prove these estimates, we note that the unitary operator~
$\Omega(s)\,:=\,U_\epsilon(s)^*\,U_n(s)$~
satisfies the differential equation
\[
\dot\Omega(s)\ =\ K(s)\ \Omega(s),\qquad\mbox{where}\qquad
K(s)\ :=\ U_\epsilon(s)^*\ [\dot P(s),\,P(s)]\ U_\epsilon(s).
\]
We cast this in its integral form
\begin{equation}\label{eq:duh}
\Omega(s)\ -\ I\ =\ \int_0^s\ K(r)\ \Omega(r)\ dr.
\end{equation}
To show that the right hand side is small, we use
the following lemmas that we prove below:

%
%

\vskip 7mm
\begin{lem}\label{lem:comm1}
~For a nearly spectral projection $P(s)$, there exist a pair~
$X(s)=X_1(s)$~ and\\ $Y(s)=Y_1(s)$,~ such that
\begin{equation}\label{eq:repr}
[\dot P(s),\,P(s)]\ =\ [X_1(s),\,H(s)]\ +\ Y_1(s),
\end{equation}
where
\begin{eqnarray}\label{eq:X}
\|X_1(s)\| &\le& \trip g_a\trip_3\ \|\dot P(s)\|,
\\[3mm]\label{eq:dotX}
\|\dot X_1(s)\| &\le& C\ \left(\,\trip g_a\trip_3\ \|\ddot P(s)\| \
+\ \trip g_a\trip_4\ \|\dot P(s)\|\,\right),
\\[3mm]
&&\hspace{-33mm}\mbox{and}\nonumber
\\[3mm]\label{eq:Y1}
\|Y_1(s)\| &\le& 2\ \delta'\ +\ 2\ \|\dot P(s)\|\ \trip g_a\trip_3\
\delta.
\end{eqnarray}
\end{lem}

%
%

\vskip 1cm
\begin{lem}\label{lem:comm2}
~For a smooth nearly spectral projection $P(s)$, there exist a pair~
$X(s)=X_2(s)$ and~ $Y(s)=Y_2(s)$,~ such that
\begin{equation}\label{eq:repr'}
[\dot P(s),\,P(s)]\ =\ [X_2(s),\,H(s)]\ +\ Y_2(s),
\end{equation}
where
\begin{eqnarray}\label{eq:X'}
\|X_2(s)\| &\le&C/a^2\,,
\\[3mm]\label{eq:dotX'}
\|\dot X_2(s)\| &\le& C\ \left(\,\trip g_a\trip_5\ \ +\ \trip
g_a\trip_4\ \|\dot P(s)\|\,\right),
\\[3mm]
&&\hspace{-33mm}\mbox{and}\nonumber
\\[3mm]\label{eq:Y2}
\|Y_2(s)\| &\le& 2\ \delta'\ +\ 2\ \|\dot P(s)\|\ \trip g_a\trip_3\
\delta \ +\ C\,\delta/a^2.
\end{eqnarray}
\end{lem}
\vspace{1cm}

\noindent
If we substitute the representation (\ref{eq:repr})
(respectively \eqref{eq:repr'}) into the right hand side of
(\ref{eq:duh}) we observe that \bea\nonumber \left\|\,\int_0^s\
K(r)\ \Omega(r)\ dr\,\right\| &\le& \int_0^s\ \|Y(r)\|\ dr
\\[3mm]\label{eq:duh_xy}
&&+\quad
\left\|\int_0^s\ U_\epsilon(r)^*\ [X(r),\,H(r)]\ U_\epsilon(r)\ \Omega(r)
\ dr\,\right\|.
\eea
However,
\[
U_\epsilon(r)^*\ [X(r),\,H(r)]\ U_\epsilon(r)\ =\
i\ \epsilon\
\frac{d\phantom{i}}{dr}\
\bigg(\,U_\epsilon(r)^*\ X(r)\ U_\epsilon(r)\,\bigg)
\ -\ i\ \epsilon\ U_\epsilon(r)^*\ \dot X(r)\ U_\epsilon(r).
\]
So, the second contribution in (\ref{eq:duh_xy}) can be integrated by
parts:
\begin{eqnarray*}
\int_0^s\ U_\epsilon(r)^*\ [X(r),\,H(r)]\ U_\epsilon(r)\ \Omega(r)\ dr
&=&
-\ i\ \epsilon\
\int_0^s\ U_\epsilon(r)^*\ \dot X(r)\ U_\epsilon(r)\ \Omega(r)\ dr
\\[3mm]
&&+\quad
i\ \epsilon\ U_\epsilon(r)^*\ X(r)\ U_\epsilon(r)\ \Omega(r)\,\Big|_0^s
\\[3mm]
&&-\quad i\ \epsilon\ \int_0^s\ U_\epsilon(r)^*\ X(r)\ U_\epsilon(r)\
\dot\Omega(r)\ dr.
\end{eqnarray*}
The norm of the first term on the right hand side is bounded by~
$\eps\,\max_s \|\dot X(s)\|$.~ The norm of the second term is
bounded by~ $2\,\eps\,\max_s \left\| X(s)\right\|$.~ Finally, since
$\Omega(r)$ is unitary, $\dot\Omega(r)\,=\,K(r)\,\Omega(r)$, and
$\|K(r)\|\,\le\,2\,\|\dot P(r)\|$, we can bound the last term by~
$2\ \eps\,\max_s \|X(s)\|\ \|\dot P(s)\|$.
Combining these estimates, we get \eqref{eq:compari} and
\eqref{eq:compari'}.

We now let~ $\psi_0\in\mbox{Ran}\,P(0)$~ be a unit vector and set~
$\psi_\epsilon(s)\ =\ U_\epsilon(s)\ \psi_0$.~
Then using (\ref{eq:n-adia}) and (\ref{2.16.5}), we see that
\bea\nonumber &&\left\|\,\psi_\eps(s)\ -\ P(s)\ U_n(s)\
\psi_\eps(0)\,\right\|
\\[3mm]\nonumber
&\le&
\left\|\,\psi_\eps(s)\ -\ U_n(s)\ \psi_\eps(0)\,\right\|\ +\
\left\|\,U_n(s)\ P(0)\ \psi_\eps(0)\ -\
P(s)\ U_n(s)\ \psi_\eps(0)\,\right\|
\\[3mm]\nonumber
&\le& \|\,U_n(s)\ -\ U_\eps(s)\,\|\ +\ \delta/\eps. \eea The theorem
now follows from~ (\ref{eq:compari})~ (\eqref{eq:compari'}
respectively).\hfill\ep

%
%

\vskip 7mm
\noindent
{\bf Proof of Lemmas \ref{lem:comm1} and \ref{lem:comm2}}\quad
The operator valued function $g_a(H(s))$ admits the
Helffer--Sj\"ostrand representation
\begin{equation}\label{eq:h-s}
g_a(H(s))\ =\ \int_{\mathbb C}\
(\partial_{\bar z}\widetilde g_a)(z)\ R_s(z)\ dz\,d\bar z,
\end{equation}
where~ $R_s(z):=\left(H(s)-z\right)^{-1}$,~~ $\widetilde g_a(z)$~ is
supported in the disc $|z|\le a$,~ and
\begin{equation}\label{eq:h-s_1}
\int_{\mathbb C}\ \left|\,\partial_{\bar z}\widetilde g_a(z)\,\right|
\ |\,\Im\,z\,|^{-n-1}\ dz\,d\bar z\ \le\
\trip g_a\trip_{n+2}\,.
\end{equation}
See {\it e.g.} \cite{Davies} or Appendix B of \cite{hs}.~
The norm here is
\[
\trip g_a\trip_{n+2}\ =\
C\ \sum_{k=0}^{n+2}\ \left\|g_a^{(k)}\right\|_{k-n-1},\qquad
\mbox{with}\qquad
\|f\|_l\ =\ \int_{-\infty}^\infty\ (1+x^2)^{l/2}\ |f(x)|\ dx,
\]
where the $C$ depends only on $n$.

Our first goal is to show that
(\ref{eq:cond_ns}) implies
\begin{equation}\label{eq:cond_g_1}
\left\|\,\Big(\,g_a(H(s))\,-\,I\,\Big)\  P(s)\,\right\|\ \,\le \ \,
\trip g_a\trip_{3}\ \delta/2.
\end{equation}
To prove this, we recall that we have taken $E(s)=0$ and note that
for any $z\in{\mathbb C}$, inequality (\ref{eq:cond_ns}) implies
\[
\left\|\,\left(H(s)- z\right)\ P(s)\ +\ z\ P(s)\,\right\|\ \le\
\delta/2.
\]
Thus,
\[
\left\|\  \frac{1}{z}\ P(s)\ +\ \left(H(s)\,-\, z\right)^{-1}\ P(s)\
\right\|\ \le\
\frac{\delta}{2\ |\,\Im\,z\,|^2}.
\]
%
Substituting this relation into (\ref{eq:h-s}) and using the bound
(\ref{eq:h-s_1}) for the error term, we get
\[
\left\|\ g_a(0)\ P(s)\ -\ g_a(H(s))\ P(s)\ \right\|\ \le\ \trip
g_a\trip_3\ \delta/2.
\]
Since~ $g_a(0)=1$,~ inequality (\ref{eq:cond_g_1}) follows.

We now define~ $X_1(s)$~ in \eqref{eq:repr} and~
$X_2(s)$~ in \eqref{eq:repr'}
to be
\bea\label{eq:X_expl}
X_1(s)&=&-\ \int_{\mathbb C}\ (\partial_{\bar z} \widetilde
g_a)(z)\ R_s(z)\ \dot P(s)\ R_s(z) \ dz\,d\bar z\,,
\\[4mm]\nonumber
&&\hspace{-3cm}\mbox{and}
\\[4mm]\label{eq:X_expl_1}
X_2(s)&=&\left(1- g_a(H(s))\right)\ R^2_s(0)\ \dot H(s)\ P(s) \
 + \ h.\,c.
\eea
Since
\[
\frac{d\phantom{i}}{ds}\,R_s(z)\ =\ -\ R_s(z)\ \dot H(s)\ R_s(z),
\]
it follows from (\ref{eq:h-s_1}) and Assumption \ref{assump_3} that
$X_1(s)$ satisfies the bounds (\ref{eq:X}) and (\ref{eq:dotX}).

Since
\be\label{eq:x1_bn}
\left\|\,\left(1- g_a(H(s))\right)\ R^2_s(0)\ (H(s)+i)\,\right\|\
\,\le\ \,C/a^2\,,
\ee
and
\[
\left\|\,R_s(-i)\ \dot H(s\,)\right\|\ \,\le\ \,C
\]
by Assumption \ref{assump_3},~ $X_2(s)$~ satisfies the
bound \eqref{eq:X'}.

To get
the bound \eqref{eq:dotX'} we first note that by partial fractions,
$$
\frac 1{z^2}\ R_s(z)\ =\
R_s(0)^2\ R_s(z)\ +\ \frac 1z\ R_s(0)^2\ +\ \frac 1{z^2}\ R_s(0).
$$
Second, since $(\alpha-z)^{-1}$ is the resolvent of multiplication by
the constant $\alpha$,
$$
g_a'(\alpha)\quad=\quad-\
\frac{d\phantom{i}}{d\alpha}\
\int_{\mathbb C}\
\frac{(\partial_{\bar z} \widetilde g_a)(z)}{(z-\alpha)}\
dz\,d\bar z
\quad=\quad
\int_{\mathbb C}\
\frac{(\partial_{\bar z} \widetilde g_a)(z)}{(z-\alpha)^2}\
dz\,d\bar z
$$
When $\alpha=0$, this is zero.
Using these facts and $g_a(0)=1$, we can write
\[
\left(1- g_a(H(s))\right)\ R^2_s(0) \ = \ -\
\int_{\mathbb C}\ \frac{(\partial_{\bar z} \widetilde g_a)(z)}{z^2}\
R_s(z)\ dz\,d\bar z\,.
\]
We then proceed as in the~ $X_1(s)$~ case.

To conclude the proof, we need to verify \eqref{eq:Y1}
and \eqref{eq:Y2}. However, we explicitly have
\[
[X_1(s),\ H(s)]\ =\ [\dot P(s),\ g_a(H(s))].
\]
Thus,
\begin{equation}\label{eq:Y_expl}
Y_1(s)\ =\ [\,\dot P(s),\ \left(P(s)\,-\,g_a(H(s))\right)\,].
\end{equation}
Since~ $P(s)\,+\,(1-P(s))\,=\,I$,~ estimate (\ref{eq:Y1}) follows from
(\ref{eq:cond_g_1}) and (\ref{eq:cond_g_2}).

On the other hand,
\begin{eqnarray*}
[X_2(s),\,H(s)]& =& \left\{\,\left(1- g_a(H(s))\right)\ R^2_s(0)\ \dot
H(s)\ P(s)\ H(s)\right.
\\[4mm]&&\quad -\  \left.\left(1-g_a(H(s))\right)\
R_s(0)\ \dot H(s)\ P(s)\,\right\}\quad -\quad h.\,c.
\end{eqnarray*}
From Assumption \ref{assump_1} with $P(s)$ smoothly nearly spectral
and $E(s)=0$, we have\\
$\ds\left\|\,\frac{d\phantom{i}}{ds}\,(H(s)\,P(s))\,\right\|
\,\le\,\delta/2$.~
So, Assumption \ref{assump_3}, and the bound
\eqref{eq:dotX'} imply that the norm of the first contribution is
$O\left(\delta/a^2\right)$, while the second contribution
is equal to
\begin{eqnarray*}
&-& \left(1-g_a(H(s))\right)\ R_s(0)\ H(s)\ \dot P(s) \ + \
O\left(\delta/a\right)
\\[3mm]&&\quad\ = \quad
-\ \left(1-g_a(H(s))\right)\ \dot P(s)\ + \
O\left(\delta/a\right)\,.
\end{eqnarray*}
Putting everything
together and using $a<1$, we get
\[
[\,X_2(s),\,H(s)\,]\ \,=\ \,[\,\dot P(s),\ g_a(H(s))\,]\ +
\ O\left(\delta/a^2\right)\,.
\]
So, $Y_2(s)$ defined by \eqref{eq:repr'} satisfies
\begin{equation}\label{eq:Y_expl'}
Y_2(s)\ =\ [\,\dot P(s),\ \left(P(s)\,-\,g_a(H(s)\right)\,]\ + \
O\left(\delta/a^2\right)\,.
\end{equation}
The rest of the argument is the same as for the~ $Y_1(s)$~ case.
 \hfill\ep

%
%

\vskip 1cm
\subsection{Proof of Theorem \ref{thm:sr}}\label{subsect2.2}
Existence of the propagator is a consequence of part (5) of
Assumption \ref{assump_sr}.

We begin the proof of Theorem \ref{thm:sr} by constructing a
suitable family $P(s)$. We note that one reasonable candidate for
$P(s)$ is a finite rank spectral projection $P_\Omega(s)$ for the
Dirichlet restriction $H_\Omega(s)$ (extended by zero to the whole
$\R^d$):
\begin{equation}\label{eq:P_Om}
P_\Omega(s)\ =\ -\ \frac 1{2\,\pi\,i}\ \oint_{\Gamma}\
\left(H_\Omega(s)\,-\,z\right)^{-1}\ dz,
\end{equation}
where  $\Gamma:=\{\,z\in\C:\,|z-E(s)|=\Delta/2\,\}$.

Indeed, for $x$ in the complement of $\partial\Omega(s)$ and any
vector $\phi$, we have \[\left(H(s)\ P_\Omega(s)\
\phi\right)(x)\,=\, E(s)\,\left(P_\Omega(s)\ \phi\right)(x)\] as
desired. Unfortunately, the range of $P_\Omega(s)$ is not in the
domain of $H(s)$, since it is not twice differentiable at the
boundary~ $\partial \Omega(s)$~ of the set $\Omega(s)$. So, we must
modify this family. The key estimate that will enable us to control
the errors introduced by this modification process is encoded in Lemma
\ref{lem:exp_dec} ({\it c.f.} with the related result in
\cite{Hislop}).

To define the desired family of projections, let
$\{\,\psi_i\,\}_{i=1}^n$ be an orthonormal basis for ${\rm
Range}\,P_\Omega(s)$, and define
\[
\widetilde \psi_i\ =\ (I\,-\,\chi_s)\ \psi_i,
\]
where $\chi_s$ is a smoothed characteristic function
of the set $\partial\Omega(s)$.
By this we mean that
$\chi_s$ is twice differentiable as a function of $x\in\R^d$,
and if $x\in\partial\Omega(s)$, then $\chi_s(x)=1$, and if~
${\rm dist}\left\{\,\partial\Omega(s),\,x\,\right\}\,>\,c/2$,~
then~ $\chi_s(x)=0$.~
We can assume $\chi_s$ has been chosen so that
$\|\chi_s\|_{2,2}\,\le\, c^{-2}$,
where
$\|\chi_s\|_{2,2}^2=\int\ |(1-\Delta_x)\chi_s(x)|^2\ dx$.

We define $P(s)$ to be the orthogonal projection
onto the span of $\{\,\widetilde \psi_i\,\}_{i=1}^n$.

\vskip 1cm
%
%
%
\begin{lem}\label{lem:closedness} ~Let
$$
Q^{(n)}(s)\,:=\,
\frac{d^n\phantom{i}}{ds^n}\,\left(P_\Omega(s)\,-\,P(s)\right)\,.
$$
 Then
\be\label{eq:closedness}
\left\|\, Q^{(n)}(s)\,\right\|\ \le\ C_n\ e^{-\eta/\hbar}/\Delta^n\,,
\ee
for $n=0,\,1,\,2$.
\end{lem}

\vskip 5mm \noindent
{\bf Proof}\quad Let
$R_w(s)\ = \ \Big((I\,-\,\chi_s)\,P_\Omega(s)\,(I\,-\,\chi_s)\,-\,
w\,I\Big)^{-1}$\quad and\quad
$\hat R_w(s) \ = \ \Big(P_\Omega(s)\,-\,w\,I\Big)^{-1}$,\quad
for~ $w\in\C$.~
Since~ $\sigma(P_\omega(s))=\{0,\,1\}$,~ Lemma \ref{lem:exp_dec}
shows that
$$
P_\Omega(s)\,-\,(I\,-\,\chi_s)\,P_\Omega(s)\,(I\,-\,\chi_s) \ = \
\chi_s \,P_\Omega(s)\,+\,P_\Omega(s)\,\chi_s\,-\,\chi_s
\,P_\Omega(s)\,\chi_s\ = \ O( e^{-\eta/\hbar}).
$$
Analytic perturbation theory then shows that for
sufficiently small $\hbar$ and $|w-1|=1/2$, we have
\begin{eqnarray}\label{eq:pr_prop1}
&&\hspace{-8mm}\left\|\,R_w(s)\,\right\| \ \le \ 3
\\[3mm]\label{eq:pr_prop2}
&&\hspace{-8mm}\left\|\,R_w(s)-\hat R_w(s)\,\right\|\ = \
O(e^{-\eta/\hbar})
\\[3mm]\label{eq:pr_prop3}
&&\hspace{-8mm}P(s)\ =\ -\ \frac{1}{2\pi i}\ \oint_\gamma\ R_w(s)\,dw
\qquad\mbox{and}\qquad
P_\Omega(s)\ =\ -\ \frac{1}{2\pi i}\
\oint_\gamma\ \hat R_w(s)\,dw\,,
\end{eqnarray}
where $\gamma:=\{w\in\C:\ |w-1|=1/2\}$.
Combining these bounds, we get the estimate \eqref{eq:closedness}
for the $n=0$ case.

For the $n=1$ case, we note that
\begin{eqnarray*}\nonumber
\frac{d\phantom{i}}{ds}\,\hat R_w(s)&=&
-\ \hat R_w(s)\ \dot P_\Omega(s)\ \hat R_w(s)\,;
\\[3mm]\nonumber
\frac{d\phantom{i}}{ds}\,R_w(s)&=&
-\ R_w(s)\ \frac{d\phantom{i}}{ds}\,
\left\{\,(I\,-\,\chi_s)\ P_\Omega(s)\ (I\,-\,\chi_s)\,\right\}\,
R_w(s)\,.
\end{eqnarray*}
Using \eqref{eq:pr_prop2}, Lemma \ref{lem:exp_dec}
and smoothness of $\chi_s$, we obtain that
\[
\frac{d\phantom{i}}{ds}\,\left(\hat R_w(s)\,-\,R_w(s)\right)\ = \
O\left(\,e^{-\eta/\hbar}/\Delta\,\right)\,,
\]
and the bound \eqref{eq:closedness} for
$n=1$ follows from the contour representations \eqref{eq:pr_prop3}.
The $n=2$ case is handled in the same way.\hfill\ep

\vskip 1cm
%
%
\begin{prop}
~Assume the hypotheses of Theorem \ref{thm:sr} and define $P(s)$ as
above. Then Assumption \ref{assump_1} holds with~ $\delta\ =\ C\
\hbar\ e^{-\eta/\hbar}$.
\end{prop}

\vskip 5mm \noindent
{\bf Proof}\quad  It is clear from the
definition of $P_\Omega(s)$ that~
$(1-\chi_s)\,(H(s)-E(s))\,P_\Omega(s)=0$.~ We therefore have
\[
(H(s)-E(s))\ (1-\chi_s)\ P_\Omega(s)\ = \
-\ [H(s),\,\chi_s]\ P_\Omega(s)\ =\ -\ [H(s),\,F_s]\ P_\Omega(s)\,,
\]
where $F_s=\chi_s \cdot\chi_{\Omega(s)}$ is a smooth function,
supported in $\Omega(s)$.
Lemma \ref{lem:exp_dec} then shows that
\[
\left\|\,(H(s)-E(s))\,(1-\chi_s)\,P_\Omega(s)\,(1-\chi_s)\,\right\|
\ \le \ C\ \hbar\ e^{-\eta/\hbar}\,.
\]
The result now follows from the identity
\[
P(s)\ =\ -\ \frac{1}{2\pi i}\ \oint_\gamma\ R_w(s)\,dw
\ = \ -\ \frac{1}{2\pi i}\
(1-\chi_s)\ P_\Omega(s)\ (1-\chi_s)\ \oint_\gamma\ w^{-1}\,R_w(s)\,dw\,,
\]
and the bound \eqref{eq:pr_prop1}.\hfill\ep

\vskip 1cm
Our next goal is to show that Assumption \ref{assump_2} is
fulfilled. Our proof relies on the geometric resolvent identity
(Lemma \ref{lem:resolvent_expansion} in Appendix).

%
%
\begin{prop}
~Assume the hypotheses of Theorem \ref{thm:sr}, and define $P(s)$ as
above. Then Assumption \ref{assump_2} holds with~
$\delta'\ =\ C\ \hbar\ e^{-\eta/\hbar}\ \trip g_a\trip_{4}/\Delta$,~
for~ $a<\Delta/2$.
\end{prop}

\vskip 5mm \noindent
{\bf Proof}\quad
We first observe that if $a<\Delta$, then $g_a(H_\Omega(s)-E(s))$
coincides with $P_\Omega(s)$. Since
\[
P_\Omega(s)\ (1\,-\,P(s))\ \dot P(s) \ = \
\left(P_\Omega(s)-P(s)\right)\ (1\,-\,P(s))\ \dot P(s)\  = \
O\left(e^{-\eta/\hbar}\right)\,,
\]
by Lemma \ref{lem:closedness}, we only need to verify that
\begin{multline}\label{eq:interi}
\left(g_a(H(s)-E(s))-g_a(H_\Omega(s)-E(s))\right)\
(1\,-\,P(s))\ \dot P(s) \\ =\
\left(g_a(H(s)-E(s))-g_a(H_\Omega(s)-E(s))\right)\ \dot P(s)\ P(s)
\phantom{\qquad}
\end{multline}
is exponentially small. We note that by construction
\[
\chi_{B(s)}\ \dot P(s)\ = \ 0,
\quad {\rm where }\quad
B\,:=\,\{x\in\R^d:\ \chi_s(x)=1\}\,\cap\,\Omega(s)\,.
\]
So, we can multiply $\dot P(s)$ in \eqref{eq:interi} by~
$\chi_{\Omega(s)\backslash B(s)}$~ from the left for free.
We now use Lemma \ref{lem:resolvent_expansion} with the choices~
$\Omega=\R^d$,~ $\Lambda_1=\Omega(s)\backslash B(s)$~ and~
$\Lambda= \Omega(s)$~ with $\partial_x\Theta$ supported on the set~
$K\subset\Omega(s)$~ with ${\rm dist}\,\{K,\,\Lambda_1\}\,>\,c'$.~
By taking adjoints, this yields
\begin{eqnarray}\nonumber
&&\left( H(s)- z\right)^{-1}\ \chi_{\Omega(s)\backslash B(s)}
\ -\ \left( H_\Omega(s)-z\right)^{-1}\ \chi_{\Omega(s)\backslash B(s)}
\\[3mm]\label{eq:re_e1}
&&=\quad
\left( H(s)-z\right)^{-1}\ [H(s),\,\Theta]\
\left( H_\Omega(s)- z\right)^{-1}\ \chi_{\Omega(s)\backslash B(s)}\,.
\end{eqnarray}
Hence
\bea\nonumber
&&\left\|\,\left\{\left(H(s)- z\right)^{-1}\
\chi_{\Omega(s)\backslash B(s)}\,-\,\left( H_\Omega(s)-z\right)^{-1}\
\chi_{\Omega(s)\backslash B(s)}\right\}\ \dot P(s)\,\right\|
\\[3mm]\nonumber
&=&
 \left\|\,\left(H(s)-z\right)^{-1}\ [H(s),\,\Theta]\
 \left( H_\Omega(s)- z\right)^{-1}\
\chi_{\Omega(s)\backslash B(s)}\ \dot P(s)\,\right\|\,,
\\[3mm]\label{eq:re_e2}
&\le&
\left\|\,\left(H(s)-z\right)^{-1}\ [H(s),\,\Theta]\,\right\|\ \, 
\left\|\,\chi_{B(s)}\left(H_\Omega(s)- z\right)^{-1}\ \dot P(s)\,\right\|\,,
\eea
where in the last step we have used
$[H(s),\,\Theta]=[H(s),\,\Theta]\,\chi_{B(s)}$ and
$\chi_{\Omega(s)\backslash B(s)}\ \dot P(s)=\dot P(s)$.
We bound the first norm by $C\,\hbar\,|\Im z|^{-1}$ with $C$
that depends only on $|z|$, using Lemma \ref{lem:a-pr}.
To estimate the second norm, we bound
\bea\nonumber
&&\left\|\,\chi_{B(s)}\ \left( H_\Omega(s)- z\right)^{-1}\
\dot P(s)\,\right\|
\\[3mm]\nonumber
&\le&\left\|\,\chi_{B(s)}\ \left( H_\Omega(s)-z\right)^{-1}\
\dot P_\Omega(s)\,\right\|\ +\
\left\|\,\chi_{B(s)}\ \left(H_\Omega(s)- z\right)^{-1}\
\left(\dot P_\Omega(s)-\dot P(s)\right)\,\right\|\,.
\eea
By Lemma \ref{lem:closedness}, the second contribution is bounded by~
$C\,e^{-\eta/\hbar}/|\Delta|$~ for~ $|z-E(s)|<\Delta/2$.
To estimate the first term, we compute
\bea\nonumber
\left( H_\Omega(s)-z\right)^{-1}\ \dot P_\Omega(s)
&=&\frac{d\phantom{i}}{ds}
\left\{\left( H_\Omega(s)- z\right)^{-1}\ P_\Omega(s)\right\}
\,-\,\frac{d\phantom{i}}{ds}
\left\{\left( H_\Omega(s)-z\right)^{-1}\right\}\ P_\Omega(s)
\\[3mm]\nonumber
&=&\frac{d\phantom{i}}{ds}
\left\{(E(s)-z)^{-1}\right\}\ P_\Omega(s)\,+\,
(E(s)-z)^{-1}\ \dot P_\Omega(s)
\\[3mm]\nonumber
&&+\quad (E(s)-z)^{-1}\,\left( H_\Omega(s)- z\right)^{-1}\
\dot H_\Omega(s)\ P_\Omega(s)\,.
\eea
Note that~ $\dot H_\Omega(s)=\dot V(s)$~ by Assumption
\ref{assump_sr}, with~ ${\rm supp}\,\dot V(s)\subset I(s)$.~
We can estimate
\bea\nonumber
&&\left\|\,\chi_{B(s)}\ \left( H_\Omega(s)-z\right)^{-1}\
\dot P_\Omega(s)\,\right\|
\\[3mm]\label{eq:lasin}
&\le&
\frac{|\dot E(s)|}{|E(s)-z|^2}\
\left\|\,\chi_{B(s)}\,P_\Omega(s)\,\right\|
\ +\ \frac{1}{|E(s)-z|}\
\left\|\chi_{B(s)}\,\dot P_\Omega(s)\,\right\|
\\[3mm]\nonumber
&&+\quad \frac{1}{|E(s)-z|}\
\left\|\,\chi_{B(s)}\ \left(H_\Omega(s)-z\right)^{-1}\ \chi_I\,\right\|
\ \,\left\|\,\dot V(s)\ P_\Omega(s)\,\right\|.
\eea 
Assumption \ref{assump_sr}, and the first resolvent formula show that
\[
\left\|\,\dot V(s)\, \left(H_\Omega(s)-z\right)^{-1}\,\right\|\
\le\ C/\Delta\qquad\mbox{when}\qquad |z-E(s)|=\Delta/2.
\]
So, using the contour representation \eqref{eq:P_Om}, we can bound~
$\left\|\,\dot V(s)\ P_\Omega(s)\,\right\|\,\le\,C$.~
All other terms in \eqref{eq:lasin} can be bounded using
Lemma \ref{lem:exp_dec} (for $|z-E(s)|<\Delta/2$).
We thus see that
\[
\left\|\,\chi_{B(s)}\
\left( H_\Omega(s)-z\right)^{-1}\ \dot P_\Omega(s)\,\right\|\ \le \
C\ \frac{e^{-\eta/\hbar}}{\Delta\ |\Im z|^2}\,.
\]
Putting everything together in \eqref{eq:re_e2}, we obtain
\bea\nonumber
&&\left\|\,\left\{\left( H(s)- z\right)^{-1}\
\chi_{\Omega(s)\backslash B(s)}\,-\,\left( H_\Omega(s)-z\right)^{-1}\
\chi_{\Omega(s)\backslash B(s)}\right\}\ \dot P(s)\,\right\|
\\[3mm]\label{eq:re_e3}
&&\le\quad C\ \frac{ \hbar\ e^{-\eta/\hbar}}{|\Im z|^3\ \Delta}\,.
\eea
The lemma now follows from the Helffer-Sj\"ostrand
representation (\ref{eq:h-s}). \hfill\ep

%
%
\section{Appendix}
\setcounter{equation}{0}

Here we collect a number of the technical statements used throughout
the text. Many of these are well-known results
that we have generalized to include magnetic fields and/or restricted
domains.

Let~ $H:=\p_\A\cdot \p_\A\,+\,V$ on $\R^d$,~
where~ $\p_\A = -i\,\hbar\,{\bf\nabla} - \A(q)$.~
Let $H_\Omega$ denote its Dirichlet restriction to the
set $\Omega$. We assume~ $H$~ is self-adjoint and that~ $H_\Omega$~
is bounded from below, so that it admits the Friedrichs extension.
Let $\Theta$ and $\tilde\Theta$ be a pair of smoothed characteristic
functions supported inside $\Omega$, and taking values in $[0,\,1]$,
such that $\tilde \Theta\,\Theta=\Theta$
(which means that $\tilde \Theta$ is ``fatter" than $\Theta$).
Throughout this Appendix,
we assume that~ $\A$~ and ~$V$~ satisfy the following hypotheses:
\begin{assump}\label{assump_tech}
\begin{enumerate}
\item[ ]
\item[{$\bf\mathcal{A}1$}]
~The components of the vector potential~ $\A$~ and their first
derivatives~ $\partial_i\A$~ for~ $i=1,...,d$~ are bounded on $\R^d$.
\item[{$\bf\mathcal{A}2$}]
~For $x$ in the support of~ ${\bf\nabla}\Theta$,~
$|V(x)|\le C$.
\end{enumerate}
\end{assump}

\vskip 5mm \noindent
We frequently estimate the norm of an operator $A$ by looking
at $A^*A$. The following bound will be used throughout the Appendix:

%
%
\vskip 4mm
\begin{lem}\label{lem:p_a}
~Let~ $H_0\,:=\,\p_\A\cdot\p_\A\,+\,1$~ and let~ $H_0^\#$~ be either~
$H_0$~ or the Friedrichs extension of its restriction to~ $\Omega$.~
Similarly, let~ $\p^\#_i$~ denote either $i$-th component of~
$\p_\A$~ or its restriction to~ $\Omega$~ (which is not self-adjoint).
If~ $G$~ is a smooth bounded function with support inside~ $\Omega$,~
then for~ $\hbar\le 1$,~ we have
\be\label{eq:p_a}
\left\|\,\p_i^\#\ G\
\left(H_0^\#\right)^{-1/2}\,\right\|\ \le\
\|\,G\,\|_\infty\,+\,C\,\hbar\,\left\|\,G\,\right\|_{2,\infty}\,.
\ee
\end{lem}

\vskip 5mm\noindent
{\bf Remark}\quad
Note that in the lemma,~ $\p_i^\Omega\ G\ =\ \p_i\ G$.

\vskip 5mm\noindent
{\bf Proof}\quad
The non-trivial part of (\ref{eq:p_a})
is the bound for the $\Omega$ restriction.
(See {\it e.g.}, \cite{simon} for the $\R^d$ case).
The first step is to show that
for any $C^1$ function $\zeta$ supported inside $\Omega$,
\be\label{1}
\left\|\,\p_i^\Omega\ \zeta\left(H^\Omega_0+t^2\right)^{-1}\,\right\|
\ \le\ \frac{4\,\left\|\,\zeta\,\right\|_{1,\infty}}{|t|\,+\,1}
\ee
for any value of the parameter~ $t\in\R$.~
To prove this, we write
\bea\nonumber
&&\left(H^\Omega_0+t^2\right)^{-1}\ \zeta\ \p_\A\cdot \p_\A\ \zeta
\left(H^\Omega_0+t^2\right)^{-1}
\\[3mm]\nonumber
&=&
\left(H^\Omega_0+t^2\right)^{-1}\ \p_\A\cdot \p_\A\ \zeta^2
\left(H^\Omega_0+t^2\right)^{-1}
\\[3mm]\label{DOG}
&&+\quad
\left(H^\Omega_0+t^2\right)^{-1}\ [\zeta,\,\p_\A\cdot \p_\A]\
\zeta \left(H^\Omega_0+t^2\right)^{-1}
\eea
The first term here is bounded by~
$\|\zeta\|_\infty^2\,(1+t^{2})^{-1}$.~
To bound the second term, note that
\[
[\zeta,\,\p_\A\cdot \p_\A]\ \zeta\ =\
[\zeta^2,\,\p_\A\cdot \p_\A]\ + \
[\,[\zeta,\,\p_\A\cdot \p_\A],\,\zeta]\,.
\]
The second contribution,~
$[\,[\zeta,\,\p_\A\cdot \p_\A],\,\zeta]$,~
is equal to~
$2\,\hbar^2\,\left({\bf \nabla}\zeta\right)^2$.~
We use this to see that by making an error whose norm is bounded
by~ $2\,\hbar^2\,\left\|\zeta\right\|^2_{1,\infty}/(1+t^2)^2$,~
the second term in (\ref{DOG}), which equals
$$
\left(H^\Omega_0+t^2\right)^{-1}\ [\zeta,\,H^\Omega_0]\
\zeta\ \left(H^\Omega_0+t^2\right)^{-1},
$$
can be replaced by
$$
\left(H^\Omega_0+t^2\right)^{-1}\ [\zeta^2,\,H^\Omega_0]\
\left(H^\Omega_0+t^2\right)^{-1}.
$$
However,
$$
\left(H^\Omega_0+t^2\right)^{-1}\
[\zeta^2,\,H^\Omega_0]\ \left(H^\Omega_0+t^2\right)^{-1}
\ \,=\ \,
\left(H^\Omega_0+t^2\right)^{-1}\ \zeta^2\ -\ \zeta^2\
\left(H^\Omega_0+t^2\right)^{-1}
$$
has norm bounded by~
$(1+t^2)^{-1}$.~
Putting the various pieces together yields \eqref{1} by
some simple estimates and~ $\hbar\le 1$.


Now we derive the bound (\ref{eq:p_a}).
First, we have
\bea\nonumber
\left\|\,\p_i^\Omega\ G\ \left(H^\Omega_0\right)^{-1/2}\,\right\|^2
&\le&
\left\|\,\left(H^\Omega_0\right)^{-1/2}\ G\ \p_\A\cdot \p_\A\
G\ \left(H^\Omega_0\right)^{-1/2}\,\right\|
\\[3mm]\nonumber
&\le&
\left\|\,\left(H^\Omega_0\right)^{1/2}\
G\ \left(H^\Omega_0\right)^{-1/2}\,\right\|^2
\eea
We use the representation
\[
\left(H^\Omega_0\right)^{-1/2}\ = \ \frac{1}{\pi}\
\int\ \left(H^\Omega_0+t^2\right)^{-1}\ dt\,,
\]
to see that
\bea\label{3}
\left(H^\Omega_0\right)^{1/2}\ G\
\left(H^\Omega_0\right)^{-1/2}\
&=&G\ +\ \left(H^\Omega_0\right)^{1/2}\
\left[G,\,\left(H^\Omega_0\right)^{-1/2}\right]
\\[3mm]\nonumber
&=&G\ +\
\frac{1}{\pi}\,\left(H^\Omega_0\right)^{1/2}\int\,
\left(H^\Omega_0+t^2\right)^{-1}[H^\Omega_0,\,G]
\left(H^\Omega_0+t^2\right)^{-1}\,dt.
\eea
Since
$$
[H^\Omega_0,\,G]\ =\
-\,2\ i\ \hbar\ {\p_\A^\Omega}\cdot({\bf\nabla}\,G)\ +\
\hbar^2\ (\Delta\,G)\,,
$$
we can use the bound \eqref{1} with~ $\zeta=\partial_jG$~
to estimate
$$
\left\|\,[H^\Omega_0,\,G]\
\left(H^\Omega_0+t^2\right)^{-1}\,\right\| \ \le \
\frac{C\,\hbar\,\left\|\,G\right\|_{2,\infty}}{|t|\,+\,1}\,.
$$
On the other hand,
$$
\left\|\,\left(H^\Omega_0\right)^{1/2}\
\left(H^\Omega_0+t^2\right)^{-1}\,\right\|\ \le\
\frac{2}{|t|+1}\,,
$$
so the right hand side of \eqref{3} is
bounded in norm by~
$\|\,G\,\|_\infty\,+\,C\,\hbar\,\left\|\,G\,\right\|_{2,\infty}$.
\hfill\ep

\vskip 5mm \noindent
Our next step is to establish the following uniform bounds:

%
%
\vskip 4mm
\begin{lem}\label{lem:a-pr}
~For the setup above, we have
\be\label{eq:res_est'}
\left\|\,[H,\,\Theta]\ \left(H-z\right)^{-1}\,\right\|\
\,\le\ \,C\,\hbar\,
\left(\,1\ +\ \frac{|z|}{\left|\Im\,z\right|}\,\right)\,,
\ee
with $C$ depending only on $\Theta$.
Furthermore, if~
$E\in\R$~ satisfies~
$\dist\left(\sigma(H_\Omega),\,E\right)\ge\Delta>0$,
then
\be\label{eq:res_est}
\left\|\,[H_\Omega,\,\Theta]\
\left( H_\Omega-z\right)^{-1}\,\right\|\ \,\le\ \,
C\,\hbar\,\left(\,1\ +\ \frac{1\,+\,|E|}{\Delta}\,\right)\,,
\ee
for any $z\in\C$, such that~ $|z-E|<\Delta/2$.~
Here $C$ depends again only on $\Theta$.
\end{lem}

\vskip 5mm\noindent
{\bf Remark}\quad
Note that if $\psi$ is smooth with support in $\Omega$,
then $[H,\,\Theta]\,\psi\,=\,[H_\Omega,\,\Theta]\,\psi$.

\vskip 5mm\noindent
{\bf Proof}\quad
Let $H_\#$ denote either $H$ or $H_\Omega$,
and similarly for the other operators that appear.
We observe that
\be\label{eq:commu}
[H_\#,\,\Theta]\ =\ \hbar^2\ (\Delta\,\Theta)\
-\ 2\ i\ \hbar\ {\bf \p_\A^\#}\cdot({\bf\nabla}\,\Theta).
\ee
Let $R^\#_z$ denote the resolvent in
(\ref{eq:res_est'}) (or in \ref{eq:res_est} accordingly), then
\be\label{eq:tog}
\left\|\,[H_\#,\,\Theta]\ R^\#_z\,\right\|\ \le\
\hbar^2\,\|\Delta\Theta\|_{\infty}\,\left\|\,R^\#_z\,\right\|
\ +\ 2\ \hbar\ \left\|\,{\p_\A^\#}\cdot({\bf\nabla}\,\Theta)\,
R^\#_z\,\right\|.
\ee
To bound the second term here, we write
\bea\nonumber
{\p_\A^\#}\cdot({\bf\nabla}\,\Theta)\ R^\#_z
&=&{\p_\A^\#}\ \tilde\Theta\
\left(H_0^\#\right)^{-1}\ H_0^\#\,\cdot\,({\bf\nabla}\,\Theta)\ R^\#_z
\\[3mm]\nonumber
&=&{\p_\A^\#}\ \tilde\Theta\ \left(H_0^\#\right)^{-1}\,\cdot\,
({\bf\nabla}\,\Theta)\  H_0^\#\ R^\#_z
\\[3mm]\label{eq:tw_t}
&&+\quad{\p_\A^\#}\ \tilde\Theta\ \left(H_0^\#\right)^{-1}\,\cdot\,
[{H_0^\#},\,{\bf \nabla}\,\Theta]\ R^\#_z\,,
\eea
where $H_0$ is defined as in Lemma \ref{lem:p_a} and $\tilde\Theta$
is defined as in the beginning of this Appendix.

Note now that
\be\label{eq:prodof}
H_0^\#\ R^\#_z\ =\ I\ +\
\left(\,(1+z)I\,-\,V_\#\,\right)\ R^\#_z.
\ee
Using this, Assumption \ref{assump_tech}, and Lemma \ref{lem:p_a},
we see that the first term on the right hand side of
(\ref{eq:tw_t}) is bounded by
\bea\nonumber
&&\hspace{-14mm}
\left\|\p_\A^\#\ \tilde\Theta\ \left(H_0^\#\right)^{-1}\right\|\
\left\|\,{\bf \nabla}\,\Theta\,\right\|_{1,\infty} 
\ \,+\ \,
\left\|\,\p_\A^\#\ \tilde\Theta\ \left(H_0^\#\right)^{-1}\,\right\|
\ \left\|\,({\bf\nabla}\,\Theta)
\cdot\left((1+z)I-V_\#\right)\,\right\|\ \left\|R^\#_z\right\|
\\[3mm]\label{eq:simil}
&\le&C\ +\
C\ (1+|z|)\ \left\|\,R^\#_z\,\right\|\,,
\eea
where we have absorbed the Sobolev norms of $\Theta$ into $C$.

To bound the second term in (\ref{eq:tw_t}), we write
\bea\nonumber
[\p_\A^\#\cdot\p_\A^\#,\ \partial_i\Theta]&=&
-\ \hbar^2\ (\Delta\,\partial_i\Theta)
-\ 2\ i\ \hbar\ \p_\A^\#\cdot({\bf\nabla}\,\partial_i\Theta)
\\[3mm]\nonumber
&=&-\ \hbar^2\ (\Delta\,\partial_i\Theta)
-\ 2\ i\ \hbar\ \tilde\Theta\ \p_\A^\#\ \tilde\Theta\,\cdot\,
({\bf\nabla}\,\partial_i\Theta)\,.
\eea
Using this, we see that
\bea\nonumber
&&\left\|\,\p^\#_\A\ \tilde\Theta\ \left(H_0^\#\right)^{-1}\
[H_0^\#,\,\partial_i\Theta]\,\right\|
\\[3mm]\nonumber
&\le&C\ \hbar^2\ 
\left\|\,\p^\#_\A\ \tilde\Theta\ \left(H_0^\#\right)^{-1}\,\right\|
\ +\ 2\ \hbar\
\left\|\,\p^\#_\A\ \tilde\Theta\ \left(H_0^\#\right)^{-1}\ \tilde\Theta\
\p^\#_\A\ \tilde\Theta\,\right\|\ \|\nabla\,\partial_i\Theta\|
\\[3mm]\nonumber
&\le&C\ \hbar^2\ +\ C\ \hbar\
\left\|\,\p^\#_\A\ \tilde\Theta\ \left(H_0^\#\right)^{-1/2}\,\right\|\
\left\|\,\left(H_0^\#\right)^{-1/2}\
\tilde\Theta\ \p^\#_\A\ \tilde\Theta\,\right\|
\\[3mm]\nonumber
&\le&C\ \hbar,
\eea
where we have repeatedly used Lemma \ref{lem:p_a} with $G=\tilde\Theta$
and the identity
$$
\tilde\Theta\ \left(\p^\#_\A\right)^*\ \tilde\Theta\ =\
\tilde\Theta\ \p^\#_\A\ \tilde\Theta.
$$
We can consequently bound the second term in (\ref{eq:tw_t}) by
$$
\left\|\,\p_\A^\#\ \tilde\Theta\ \left(H_0^\#\right)^{-1}\,\cdot\,
[\p_\A^\#\cdot\p_\A^\#,\ {\bf \nabla}\,\Theta]\ R^\#_z\,\right\|
\ \le\ C\ \hbar\ \left\|\,R^\#_z\,\right\|\,.
$$
Putting everything together into (\ref{eq:tog}), the bounds
(\ref{eq:res_est'}) and (\ref{eq:res_est}) follow from\\
$\ds\left\|\,R_z\,\right\|\le 1/|\Im z|$,\quad and\quad
$\left\|\,R^\Omega_z\,\right\|\le 2/\Delta$~ whenever~
$|E-z|<\Delta/2$.\hfill\ep

\vskip 5mm \noindent
Many of the technical results throughout the paper rely on the following
general result (Lemma 4.2 in \cite{aenss}) that is known as the
geometric resolvent identity.

%
%
\begin{lem}\label{lem:resolvent_expansion}
~[The Geometric Resolvent Identity]
~Let $H$ be a Schr\"odinger operator on $\R^d$.
Consider four open sets $\Lambda_1$, $\Lambda_2$, $\Lambda$, and $\Omega$
that satisfy~ $\Lambda_1\subset\Lambda$,~ $\Lambda_2\subset\Lambda$,~
$\Lambda\subset\Omega$, and~
$\mbox{dist}\{\Lambda_1\cup\Lambda_2,\,\Lambda^c\}>0$.~
Let $\Theta$ be a smooth function which is identically $1$ on a
neighborhood of~ $\Lambda_1\cup\Lambda_2$~ and identically $0$ on a
neighborhood of~ $\Lambda^c$.~
Given any
restrictions~ $H_\Omega$~ and~ $H_\Lambda$ of~ $H$ to~ $\Omega$~
and~ $\Lambda$,~ respectively, we have
\begin{eqnarray}\label{eq:re_exp}
\chi_{\Lambda_1}\ \left( H_\Omega-z\right)^{-1} &=&\chi_{\Lambda_1}\
\left( H_\Lambda-z\right)^{-1}\ \Theta
\\[3mm]\nonumber
&&+\quad \chi_{\Lambda_1}\ \left( H_\Lambda-z\right)^{-1}\
[H,\,\Theta]\ \left( H_\Omega-z\right)^{-1}
\end{eqnarray}
for any $z$ for which both resolvents exist. Also,
\begin{eqnarray}\label{eq:re_exp_1}
\chi_{\Lambda_1}\ \left( H_\Omega-z\right)^{-1}\ \chi_{\Lambda_2}
&=&\chi_{\Lambda_1}\ \left( H_\Lambda-z\right)^{-1}\
\chi_{\Lambda_2}
\\[3mm]\nonumber
&&+\quad \chi_{\Lambda_1}\ \left( H_\Lambda-z\right)^{-1}\
[H,\,\Theta]\ \left( H_\Omega-z\right)^{-1}\ \chi_{\Lambda_2}\,,
\end{eqnarray}
under the same conditions.
\end{lem}

%
%
\vskip 5mm\noindent
{\bf Proof}\quad
Since the function~ $\Theta$~
has support strictly contained within~ $\Lambda$,~ multiplication
by~ $\Theta$~ satisfies~ $H_\Lambda\ \Theta\,=\,H_\Omega\ \Theta$~
on~ ${\mathcal D}(H_{\Omega})$.~ Thus, on~ ${\mathcal D}(H_{\Omega})$,
$$
[H,\,\Theta]\ =\ (H_\Lambda-z)\ \Theta\ -\ \Theta\ (H_\Omega-z)\ .
$$
We multiply this on the left by $(H_\Lambda-z)^{-1}$ and on the
right by $(H_\Omega-z)^{-1}$ to see that
$$
\Theta\ (H_\Omega-z)^{-1}\ = \
(H_\Lambda-z)^{-1}\ \Theta\ +\
(H_\Lambda-z)^{-1}\ [H,\,\Theta]\
(H_\Omega-z)^{-1}.
$$
Multiplying both sides of  the above equation by $\chi_{\Lambda_1}$
from the left and using
$\chi_{\Lambda_1}\cdot\Theta=\chi_{\Lambda_1}$
gives \eqref{eq:re_exp}. Multiplying both sides of
\eqref{eq:re_exp} by $\chi_{\Lambda_2}$ on the right gives
\eqref{eq:re_exp_1}.\hfill \ep

\vskip 5mm
Armed with this tool, we can prove
%
%
%
%
\vskip 3mm
\begin{thm}\label{thm:CT}
~[The Combes--Thomas estimate]
~Let $H=\p_\A\cdot \p_\A\,+\,V$ be as above.
Suppose that on the domain $J$, the potential $V$ is greater than $E+b$,
for some $\hbar$--independent $b>0$. Then there exists an
$\hbar$--independent $\eta>0$, such that the resolvent of the
Dirichlet restriction $H_J$ of $H$ to $J$ satisfies
\be\label{eq:exp_dec}
\left\|\,\chi_{J_i}\,\left(H_{J}-z\right)^{-1}\,\chi_{J_j}\,\right\|\
\le\ K\,\left(\hbar^{-1}+\frac{1\,+\,|E|}{b}\right)\ e^{-\eta/\hbar}
\ee
for $z\in\C$, such that~ $|E-z|<b/2$,~ and any~ $J_{i,j}\subset J$~
that satisfy
\[
{\rm dist}\,\left\{J_i,\ J_j\right\}\ \ge\ c/2.
\]
Here $K$ is a constant that depends only on $c$.
\end{thm}

%
%
\vskip 5mm \noindent
{\bf Proof of Theorem \ref{thm:CT}}\quad
Consider the operator
$\tilde H:=\hbar^2\,\p_\A\cdot \p_\A\,+\,\tilde V(x)$
acting on $\R^d$, where
\[
\tilde V(x)\ = \
\begin{cases}
~V(x)&~\mbox{if}~ x\in J\\[1mm]
~E+b&~\mbox{otherwise.}
\end{cases}
\]
Then the (improved) Combes-Thomas estimate \cite{GK} is
applicable for $\tilde H$ and implies
\be\label{eq:exp_dec_r}
\left\|\,\chi_{J_i}\,\left(\tilde H-z\right)^{-1}\,\chi_{J_j}\,\right\|\
\le\ \tilde K\ \hbar^{-1}\ e^{-\eta/\hbar}\,.
\ee
for~ $|E-z|<b/2$~ and~ $\dist\,\{J_i(s),\,J_{j}(s)\}\ge c/8$,~
and some generic constant~ $\tilde K$.

If we now use the geometric resolvent
identity \eqref{eq:re_exp_1}, with~ $\chi_{\Lambda_i}=\chi_{J_{i}}$,~
$\Omega=\R^d$,~ and\\ $\Lambda=J$,~ where~ $\Theta$~ satisfies
\[
\dist\,\{\supp\left({\bf \nabla} \Theta\right),\,\chi_{J_{i}}\}\ \ge\
c/8\,,\quad\mbox{for}\quad i=1,\,2\,,
\]
we get
\bea\label{eq:re_exp_2}
\hspace{-1cm}
\left\|\,\chi_{J_1}\ \left(H_{J}-z\right)^{-1}\ \chi_{J_2}\,\right\|
&\le&
\left\|\,\chi_{J_1}\ \left(\tilde H-z\right)^{-1}\ \chi_{J_2}\,\right\|
\\[3mm]\nonumber
&&+\quad
\left\|\,\chi_{J_1}\ \left(H_{J}-z\right)^{-1}\ [H,\,\Theta]\,\right\|
\quad
\left\|\,\chi_{J_3}\ \left(\tilde H-z\right)^{-1}\ \chi_{J_2}\,\right\|\,,
\end{eqnarray}
where $J_3$ can be chosen in such a way that
\[
[H,\,\Theta]\ \chi_{J_3}\ =\  [H,\,\Theta]\,,
\quad\mbox{and}\quad\dist\,\{J_3,\,J_{2}\}\,\ge\,c/8\,.
\]
Using \eqref{eq:res_est} and \eqref{eq:exp_dec_r}, we bound
the right hand side of \eqref{eq:re_exp_2} by
\[
K\,\left(\,\hbar^{-1}\,+\,\frac{1\,+\,|E|}{b}\,\right)\ e^{-\eta/\hbar}\,,
\]
since the gap $\Delta$ in \eqref{eq:res_est}
in the situation at hand is $b$. \hfill\ep

\vskip 5mm \noindent
The Combes-Thomas estimate leads to the following result, where we suppress
the $s$--dependence whenever possible for the sake of brevity.

%
%
%
\begin{lem}\label{lem:exp_dec}
~[Exponential bounds for the spectral projection]
~Let $H_\Omega$ satisfy Assumption \ref{assump_sr}.
Suppose $F$ and its first and second derivatives are smooth and bounded,
and that $F$ vanishes on~
$B\ :=\ \{x\in\R^d:\ \dist\,\{x,\,\partial\Omega\}>c/8\}$.~
Then
\bea\label{eq:falloff}
\left\|\,F\ P_\Omega\,\right\|&\le&C\ e^{-\eta/\hbar},
\\[3mm]\label{eq:falloff2}
\left\|\,F\ \dot P_\Omega(s)\,\right\|&\le& C\
e^{-\eta/\hbar}/\Delta\,,\qquad\
\left\|\,F\ \ddot{P}_\Omega(s)\,\right\|\ \le\ C\
e^{-\eta/\hbar}/\Delta^2\,,
\\[3mm]\label{eq:falloff3}
\left\|[H_\Omega,\,F]\, P_\Omega(s)\,\right\|&\le&
C\ \hbar\ e^{-\eta/\hbar}\,,
\\[3mm]\nonumber
&&\hspace{-58mm}\mbox{and}
\\[3mm]\label{eq:falloff1}
\left\|\,F\ \left( H_\Omega - z\right)^{-1}\ \chi_{\widetilde
I}\,\right\| &\le&C\ e^{-\eta/\hbar}/\Delta,
\quad~\mbox{for}~\quad |z-E|\,<\,\Delta/2.
\eea
Here $\widetilde I$ is a set that is slightly larger than $I$ and
satisfies~ $I\subset\widetilde I$~ and
%
\[
{\rm dist}\,\{I,\,\widetilde I^c\,\}\ \ge\ c'\,,\qquad
{\rm dist}\,\{B,\,\widetilde I^c\,\}\ \ge\ 3\,c/4
\]
for some $\hbar$--independent~ $c'>0$.
The constant $C$ depends only on $c$, $c'$ and $E$.
\end{lem}
%

%
%
\vskip 5mm\noindent
{\bf Proof of Lemma \ref{lem:exp_dec}}\quad
Assertions \eqref{eq:falloff} -- \eqref{eq:falloff2} are consequences
of (\ref{eq:falloff1}), so we prove (\ref{eq:falloff1}) first.
To do so, we make use of Lemma \ref{lem:resolvent_expansion}.
We let $\Lambda=\Omega\bigcap J$ and $\Lambda_o\subset\Lambda$ be a
neighborhood of $B$.~
We choose $\Theta$ so that $\partial_x\Theta$ is
supported in a set $S\subset J\backslash\widetilde I$, such that
${\rm dist}\left\{S,\,B\right\}\ge c/2$.
Lemma \ref{lem:resolvent_expansion} then yields
\be\label{eq:re_exp1}
F\ \left( H_\Omega-z\right)^{-1}\ \chi_{\tilde I}\ =\
F\left(H_\Lambda-z\right)^{-1}\ [H,\,\Theta]\
\left( H_\Omega-z\right)^{-1}\ \chi_{\widetilde I}.
\ee
Since~ $\chi_{S}\ [H,\,\Theta]\ = \ [H,\,\Theta] $,~ we obtain
\begin{eqnarray}\label{eq:re_exp2}
F\ \left( H_\Omega-z\right)^{-1}\ \chi_{\widetilde I}&=&
\left(F\ \left( H_\Lambda-z\right)^{-1}\ \chi_S\right)\
\left(\,[H,\,\Theta]\ \left(H_\Omega-z\right)^{-1}\
\chi_{\widetilde I}\,\right).
\end{eqnarray}
Note that for  $|z-E|<\Delta/2$, the first term in the parentheses on
the right hand side enjoys the property (\ref{eq:exp_dec}). Putting
together (\ref{eq:re_exp2}), (\ref{eq:exp_dec}), and
(\ref{eq:res_est}), we get (\ref{eq:falloff1}).

Note now that\quad $\chi_{\tilde I}\ \dot H(s)\ =\ \dot H(s)$,
$$
\frac{d\phantom{i}}{ds}\,\left(H_\Omega(s)-z\right)^{-1}\ = \
-\ \left(H_\Omega(s)-z\right)^{-1}\,\dot H(s)\
\left(H_\Omega(s)-z\right)^{-1}
$$
and
$$
\left\|\dot H(s)\,\left( H_\Omega(s)-z\right)^{-1}\right\|\ \le \
C/\Delta\qquad
{\rm for }\qquad |z-E(s)|=\Delta/2
$$
by Assumption \ref{assump_sr}.
It follows from the integral representation (\ref{eq:P_Om}) and the
bound (\ref{eq:falloff1}) that the first assertion in
\eqref{eq:falloff2} holds. The second assertion in
\eqref{eq:falloff2} is established in the same way by taking the
second derivative of the resolvent
(with respect to the $s$ variable).

We now prove the bound \eqref{eq:falloff}. The integral representation
(\ref{eq:P_Om}) and the bound (\ref{eq:falloff1}) show that
$$
\left\|\,F\ P_\Omega\ \chi_{\tilde I}\,\right\|\ \le\ C\
 e^{-\eta/\hbar}.
$$
The desired bound (\ref{eq:falloff}) will be obtained if we can show
that
$$
\left\|\,F\  P_\Omega\ \chi_{\tilde I}\,\right\|\ \le\
C\ e^{-\eta/\hbar}\qquad\mbox{implies}\qquad \left\|\,F\ P_\Omega\,
\right\|\ \le\ \tilde C\ e^{-\eta/\hbar}\,.
$$
To prove this assertion, it suffices to show that for every~
$\psi\in{\rm Range}\,P_\Omega$~ with~ $\|\psi\|=1$,~ we have
\be\label{eq:pos_o}
\|\,\chi_{\tilde I}\,\psi\,\|\ \ge\ K\ >\ 0\,.
\ee
Indeed,  it follows from \eqref{eq:pos_o} that
\[
P_\Omega\ \chi^2_{\tilde I}\ P_\Omega \ > \ K^2\, P_\Omega\,,
\]
so
\[
F\  P_\Omega\ \chi^2_{\tilde I}\ P_\Omega\ F\ >\
K^2\,\left(F\ P_\Omega\ F\right)\,,
\]
and the result follows.

Inequality \eqref{eq:pos_o} is a consequence of the bound
\begin{equation}\label{eq:bnd}
\|\,F_{I}\,\psi\,\|\ \ge\ K\,,
\end{equation}
for some smoothed characteristic function~ $F_{I}$~ of the set~ $I$~
that satisfies
\[
F_{I}\ \chi_{\tilde I}\ =\ F_{I}\,, \qquad\mbox{and}\qquad F_I(x)=1\quad
{\rm for}\quad x\in I\,.
\]
To show that (\ref{eq:bnd}) holds, let
$$
\psi_1\ :=\ F_{I}\ \psi,\qquad \psi_2\ :=\ \left(I\ -\
F_{I}\right)\,\psi,
\qquad\mbox{and}\qquad
\psi\ =\ \psi_1\ +\ \psi_2\,.
$$
Then
\[
\langle\,\psi_2,\ H_\Omega\ \psi_2\,\rangle\ \ge\
(E+b)\ \|\psi_2\|^2
\]
by the definition of the region $J$.~ On the other hand,
\begin{eqnarray*}
E&=&\langle\,\psi,\ H_\Omega\ \psi\,\rangle
\\[3mm]
&=&\langle\,\psi_1,\ H_\Omega\ \psi_1\,\rangle\ +\ \langle\,\psi_2,\
H_\Omega\ \psi_2\,\rangle\ +\ 2\ \Re\ \langle\,\psi_1,\ H_\Omega\
\psi_2\,\rangle
\\[3mm]
&\le&E\ \|\psi_1\|^2\,+\,(E+b)\ \|\psi_2\|^2\,-\ E\,\
\langle\,\psi_1,\,\psi_2\rangle\,+\,
\langle\,[H_\Omega,\,F_I]\ \psi,\,\psi_1\,\rangle
\\[3mm]
&=&E\ +\ b\ \|\psi_2\|^2\ +\
\langle\,[H_\Omega,\,F_I]\ \psi,\,\psi_1\,\rangle.
\end{eqnarray*}

We now use the integral representation (\ref{eq:P_Om}) together with
the bound (\ref{eq:res_est}) to see that
\[
\left\|\,[H,\,F_I]\ \psi\,\right\|\ \le \ C\ \hbar\
\|\,F_I\,\|_{2,\infty}\,,
\]
where $C$ is independent of the choice of~
$\psi\in {\rm Range}\,P_\Omega$.~
So, with the appropriate choice of $F_I$,~ we see that~
$\|\,\psi_2\,\|^2\ \le\ C\ \hbar$,~ and (\ref{eq:bnd}) follows.

Finally, we show that \eqref{eq:falloff3} follows from
\eqref{eq:falloff}. The proof closely follows the proof of
Lemma \ref{lem:p_a}.
We use \eqref{eq:commu} and \eqref{eq:falloff} to bound
\[
\left\|\,[H_\Omega,\,F]\ P_\Omega\,\right\|\ \le\
\,2\ \hbar\ \left\|\,\p_\A\ \tilde F\,P_\Omega\,\right\|\ +\
C\ \hbar^2\,e^{-\eta/\hbar},
\]
where $\tilde F:=|{\bf\nabla}\,F|$.
Now, we have
\be\label{eq:bndpo}
P_\Omega\ \tilde F\ \p_\A\cdot \p_\A\ \tilde F\ P_\Omega\ = \
P_\Omega\ \tilde F^2\ \p_\A\cdot \p_\A\  P_\Omega \ +\
P_\Omega\ [[\tilde F,\,\p_\A\cdot \p_\A],\,\tilde F]\ P_\Omega\,.
\ee
The first term is bounded by
\[
\left\|\,P_\Omega\ \tilde F\,\right\|\quad
\left\|\,\tilde F\ H_0^\Omega\  P_\Omega\,\right\|\ \le \
C\,e^{-\eta/\hbar}\ ,
\]
using the bound \eqref{eq:falloff}, the identity \eqref{eq:prodof}
with the estimate in \eqref{eq:simil}, as well as the contour
representation \eqref{eq:P_Om}.
Since
$\left[\,(\tilde F,\p_\A\cdot \p_\A),\,\tilde F\,\right]\ = \
2\,\hbar^2\,\left({\bf \nabla}\Theta\right)^2$,
the second term in \eqref{eq:bndpo} is
bounded by~ $C\,e^{-\eta/\hbar}$~ using \eqref{eq:falloff}.
\hfill\ep

\vskip 1cm \noindent
{\Large\bf Acknowledgement.}\quad It is a pleasure to thank
Professor Gian Michele Graf for several useful discussions.

\phantom{ }

\end{document}